\documentclass[twocolumn,showpacs,nofootinbib]{revtex4}

\usepackage{cancel}
\usepackage{graphicx}
\usepackage{epsfig}
\usepackage{amsmath,amsfonts,amssymb}
\usepackage[colorlinks=false]{hyperref}
\usepackage{multirow}
\def\al{\alpha}   
   
   \def\ka{\kappa}

 \def\Om{\Omega}

 \def\frac#1#2{{\textstyle{{#1}\over
{#2}}}} 
\def\lsim{\mathrel{\rlap{\lower4pt\hbox{\hskip1pt$\sim$}}
\raise1pt\hbox{$<$}}}
\def\gsim{\mathrel{\rlap{\lower4pt\hbox{\hskip1pt$\sim$}}
\raise1pt\hbox{$>$}}} \def\sqr#1#2{{\vcenter{\vbox{\hrule height.#2pt
\hbox{\vrule width.#2pt height#1pt \kern#1pt \vrule width.#2pt} \hrule
height.#2pt}}}}

\def\beq{\begin{equation}} \def\eeq{\end{equation}}
\def\beqa{\begin{eqnarray}} \def\eeqa{\end{eqnarray}}
\def\eq#1{Eq. (\ref{#1})}
\begin{document}

\title{Dynamical analysis of nonminimal coupled theories}

\author{Rafael Ribeiro}
\email{rafael.ribeiro@tecnico.ulisboa.pt}
\affiliation{Departamento de F\'{\i}sica,\\ Instituto Superior T\'ecnico, Universidade de Lisboa, \\Av. Rovisco Pais 1, 1049-001 Lisboa, Portugal}

\author{Jorge P\'aramos}
\email{jorge.paramos@fc.up.pt}
\affiliation{Departamento de F\'isica e Astronomia and Centro de F\'isica do Porto, Faculdade de Ci\^encias da Universidade do Porto, Rua do Campo Alegre 687, 4169-007 Porto, Portugal}

\date{\today}

\begin{abstract}
In this work a dynamical system approach to nonminimal coupled $f(R)$ theories is made. The solutions of three distinct models are obtained and their stability and physical interpretation are studied to ascertain their viability as candidates for dark energy. Comparison is drawn with previous works in the context of $f(R)$ and nonminimally coupled models.
\end{abstract}
\maketitle
\section{Introduction}\label{section:introduction}
Despite its experimental success \cite{experimental}, it is known that general relativity (GR) does not exhibit the most general form to couple matter with curvature. In fact, these can be coupled in a nonminimal way \cite{Bertolami:2007gv} (for early proposals see Ref. \cite{early}) that has already been shown to be able to mimic dark matter \cite{NMCDM}, dark energy \cite{Bertolami:2010cw,Bertolami:2011fz,Bertolami:2013uwl} and explain post-inflationary preheating \cite{Bertolami:2010ke} and cosmological structure formation \cite{NMCS}.

This nonminimal coupling (NMC) can give rise to several implications, from Solar System \cite{Bertolami:2013qaa} and stellar dynamics \cite{NMCstellar} to close like-time curves \cite{Bertolami:2012fz}, wormholes \cite{MontelongoGarcia:2010xd}, black holes \cite{NMCbh}, modifications to virial equilibrium \cite{virial}, Palatini formulation \cite{Palatini}, absence of Dolgov-Kawasaki instabilities and the well-known energy conditions \cite{NMCDK} (see Ref. \cite{Bertolami:2013xda} for a thorough review). Previous proposals to address dark energy \cite{singleNMCDE,repetido} or inflation (chaotic \cite{singleNMCCI} or Higgs-induced \cite{singleNMCHI}) had included a NMC between the scalar curvature and a scalar field, but did not extend this coupling to the baryonic matter content.

From a fundamental standpoint, a NMC can arise from one-loop vacuum-polarization effects in the formulation of quantum electrodynamics in a curved spacetime \cite{Drummond:1979pp}, as well as in the context of matter scalar fields \cite{Damour:1992we,Bertolami:2008im}. In the framework of Riemann-Cartan geometry, a NMC was considered in an earlier proposal \cite{Goenner:1984zx} and another study showed that it clearly affects the features of the ground state \cite{Bertolami:1987wj}. Phenomenologically, it can be viewed as a natural continuation of so-called $f(R)$ theories \cite{felice,fR}, where the standard Einstein-Hilbert action is replaced by a non-linear function $f(R)$ of the scalar curvature --- an extension of GR that has garnered a strong interest in the past decade.

Other theories were proposed to address the cosmological problems, like quintessence \cite{quintessence,repetido} and Gauss-Bonnet models \cite{GB}. An unification of dark components was also suggested using a generalized Chaplygin gas \cite{GCG}.

The purpose of this work is to make a dynamical system approach on NMC theories in a cosmological setting and derive the solutions for some models; for a very recent and similar study, albeit less general, see Ref. \cite{Azizi:2014qsa}. It is similar in scope to other studies in the context of $f(R)$ \cite{Carloni:2004kp,Carloni:2007br} and $f(T,T_G)$ theories \cite{fttg}.

This work is organized as follows: the nonminimal gravitational model is discussed in Sec. \ref{section:model}; the formulation of the equivalent dynamical system is presented in Sec. \ref{section:dynamical}; a confirmation of the dynamical system obtained in $f(R)$ theories is shown in Sec. \ref{section:fR}; the discussions of the results obtained for two pure NMC models and for a power law correction model are presented in Secs. \ref{section:pureNMC} and \ref{section:powerlaw}, respectively. Finally, the conclusions are presented in Sec. \ref{section:conclusion}.
\section{The Model}\label{section:model}
Following the generalization of the Einstein-Hilbert action put forward in $f(R)$ theories \cite{felice}, a NMC model is embodied in the action \cite{Bertolami:2007gv},
\begin{equation}
S=\int d^4x\sqrt{-g}\left[\kappa f_1(R) + f_2(R)\mathcal{L}\right],
\end{equation}
where $\kappa=c^4/(16\pi G)$, $f_i(R)$ are arbitrary functions of the scalar curvature $R$, $g$ is the metric determinant and $\mathcal{L}$ is the matter Lagrangian density; the standard Einstein-Hilbert action is obtained by taking $f_1(R)=R-2\Lambda$ and $f_2(R)=1$. The field equations are obtained by imposing a null variation of the action with respect to the metric, 
\begin{equation}
FG_{\mu\nu}=\dfrac{1}{2}f_2T_{\mu\nu}+\bigtriangleup_{\mu\nu}F+\dfrac{1}{2}g_{\mu\nu}\kappa f_1 - \dfrac{1}{2}g_{\mu\nu}RF,
\label{feq}
\end{equation}
where $F=\kappa f'_1 + f'_2 \mathcal{L}$, the prime denotes derivation with respect to the scalar curvature (omitted), $\bigtriangleup_{\mu\nu}\equiv\bigtriangledown_{\mu}\bigtriangledown_{\nu}-g_{\mu\nu}\Box$, and the matter energy-momentum tensor is defined as
\begin{equation}
T_{\mu\nu}=-\dfrac{2}{\sqrt{-g}}\dfrac{\delta\left(\sqrt{-g}\mathcal{L}\right)}{\delta g^{\mu\nu}}.
\end{equation}
The Bianchi identities imply the non-covariant conservation law
\begin{equation}
\bigtriangledown^{\mu}T_{\mu\nu}=\dfrac{f'_2}{f_2}\left(g_{\mu\nu}\mathcal{L}-T_{\mu\nu}\right)\bigtriangledown^{\mu}R.
\label{conserv}
\end{equation}
Since there is an equivalence between this model and a two-scalar field model, this non-conservation may be interpreted as an energy exchange between matter and those scalar fields \cite{Bertolami:2008im}, and can also lead to a deviation from geodesic motion \cite{Puetzfeld}.

To study the recent accelerated expansion of our universe,  a flat universe is considered with the line element
\begin{equation}
ds^2=-dt^2+a^2(t)dV^2,
\label{metric}
\end{equation}
where $a(t)$ is the scale factor and $dV$ is the volume element with comoving coordinates, and matter is assumed to behave as a perfect fluid, with an energy-momentum tensor
\begin{equation}
T^{\mu\nu}=\left(\rho+P\right)u^{\mu}u^{\nu}+Pg^{\mu\nu},
\end{equation}
derived from the Lagrangian density $\mathcal{L}=-\rho$ (see Refs. \cite{fluid} for a discussion), where $\rho$ and $P$ are the energy density and pressure of the perfect fluid, respectively, and $u^{\mu}$ is its four-velocity. 

One can see that the energy-momentum tensor is again conserved, just like in GR or $f(R)$ theories, since Eq. (\ref{conserv}) yields the continuity equation
\begin{equation}
\dot{\rho}+3H(1+w)\rho=0,
\label{conteq}
\end{equation}
where $H=\dot{a}/a$ is the Hubble parameter and $w=P/\rho$ is the equation of state (EOS) parameter.

Inserting the metric in the field equations (\ref{feq}), one obtains the modified field equations
\begin{equation}
\label{friedmann}
H^2=\dfrac{1}{3F}\bigg[\dfrac{1}{2}FR  - 3HF'\dot{R}-\dfrac{1}{2}\kappa f_1+ \dfrac{1}{2}f_2\rho-9H^2(1+w)f_2'\rho \bigg],
\end{equation}
and
\begin{equation}
2\dot{H}+3H^2=\dfrac{1}{2F}\left[ FR -\kappa f_1 - 2\ddot{F} - 4H\dot{F} - f_2w\rho \right].
\label{racheq}
\end{equation}
Notice that \eq{racheq} can be obtained by differentiating the modified Friedmann \eq{friedmann}, as shall be evoked in the following section.

\section{Dynamical System}\label{section:dynamical}
One way to obtain the solutions of the field equations is via the study of the ensuing dynamical system, written in terms of the dimensionless variables
\begin{eqnarray}
\label{variables}
x=-\dfrac{F'\dot{R}}{FH},~~ y=\dfrac{R}{6H^2}, ~~z=-\dfrac{\kappa f_1}{6FH^2},\\ 
\Omega_1=\dfrac{f_2\rho}{6FH^2},~~~~\Omega_2=-\dfrac{3(1+w)f_2'\rho}{F},\nonumber
\end{eqnarray}
with $F'\equiv\kappa f_1''-f_2''\rho$ the partial derivative of $F$ with respect to the scalar curvature $R$). Notice that the introduction of the nonminimal coupling increases the number of variables of the problem --- for $f(R)$ theories, only four variables were required \cite{Carloni:2007br}.

The modified Friedmann equation (\ref{friedmann}) becomes
\begin{equation}
1=x+y+z+\Omega_1+\Omega_2,
\label{friedmanndyn}
\end{equation}
acting as a restriction to the phase space.

In terms of the quantities defined in \eq{variables}, one has
\begin{eqnarray}\label{Fconstant}
\dfrac{\dot{F}}{FH} &=& - ( x + \Om_2 ) \to \\ \nonumber \dfrac{\ddot{F}}{FH^2} &=& (2-y+x+\Om_2)(x+\Om_2)- \dfrac{dx}{dN}-\dfrac{d\Om_2}{dN},
\end{eqnarray}
where $N=\ln a$ is the number of e-folds. This implies that, for a constant $F$ (as studied in Ref. \cite{Bertolami:2013uwl}), the additional constraint $ x + \Om_2 = 0$ holds.

Furthermore, the Raychaudhuri \eq{racheq} becomes
\begin{equation}
\dfrac{dx}{dN} + \dfrac{d\Om_2}{dN}= (x+\Om_2) (x + \Om_2 - y) -y -3z + 3w\Omega_1 - 1.
\end{equation}
Due to the conservation law \eq{conteq}, one may directly compute the following,
\begin{equation}
\dfrac{d\Om_2}{dN} = \Omega_2\left[x\left(1-\dfrac{\al_2}{\al}\right)-3\left(1+w\right)+\Omega_2\right],
\end{equation}
so that the Raychaudhuri \eq{racheq} translates into
\begin{eqnarray}\label{translates}
\dfrac{dx}{dN} &=& x\left[x-y+\Omega_2\left(1+\dfrac{\al_2}{\al}\right)\right]-1-y-3z \\ \nonumber && + 3w\Omega_1+\Omega_2\left[3\left(1+w\right)-y\right].
\end{eqnarray}
Differentiating the remaining variables with respect to $N$, one obtains the following autonomous system, equivalent to the field equations (\ref{friedmann}),
\begin{equation}
\begin{cases}\vspace{2mm}
\dfrac{dx}{dN}=x\left[x-y+\Omega_2\left(1+\dfrac{\al_2}{\al}\right)\right]-1-y-3z+ \\\vspace{2mm}
~~~~~~~~~3w\Omega_1+\Omega_2\left[3\left(1+w\right)-y\right]\\\vspace{2mm}
\dfrac{dy}{dN}=y\left[2\left(2-y\right) -\dfrac{x}{\alpha} \right]\\\vspace{2mm}
\dfrac{dz}{dN}=z\left[x\left(1-\dfrac{\alpha_1}{\al}\right)+\Omega_2+2\left(2-y\right)\right]\\\vspace{2mm}
\dfrac{d\Omega_1}{dN}=\dfrac{\Omega_2xy}{3\al \left(1+w\right)}+\Omega_1\left(1-3w+x+\Omega_2-2y\right)\\\vspace{2mm}
\dfrac{d\Omega_2}{dN}=\Omega_2\left[x\left(1-\dfrac{\al_2}{\al}\right)-3\left(1+w\right)+\Omega_2\right]\end{cases},
\label{dynsystem}
\end{equation}
subject to the constraint \eq{friedmanndyn} and with the dimensionless parameters,
\begin{equation}
\label{defalphas}\alpha(R,\rho)=\dfrac{F'R}{F},~~~~ \al_1(R)=\dfrac{f_1'R}{f_1},~~~~ \al_2(R)=\dfrac{f_2''R}{f_2'}.
\end{equation} 
One useful relation is
\begin{equation}
\alpha=\dfrac{f_1''R}{f_1'}\left[1-\dfrac{\Omega_2}{3(1+w)}\right]+\dfrac{\al_2\Omega_2}{3(1+w)}. \label{alpha}
\end{equation}

As highlighted in the previous section, the Raychaudhuri \eq{racheq} is equivalent to the relation (\ref{translates}) for $dx/dN$. However, the former can also be computed by differentiating the Friedmann \eq{friedmann}, as noticed in the previous section: since this should always hold, one must have 
\begin{equation}
\dfrac{dx}{dN} + \dfrac{dy}{dN} + \dfrac{dz}{dN} + \dfrac{d\Omega_1}{dN} + \dfrac{d\Omega_2}{dN} = 0.
\end{equation}
However, the sum of all the equations of the system (\ref{dynsystem}) does not vanish trivially: instead, one obtains the additional constraint
\begin{equation}
\label{raychaudhuridyn}
y\left[\dfrac{\Omega_2}{3(1+w)}-1\right] = z \al_1 ,
\end{equation}
a direct consequence of the Raychaudhuri \eq{racheq}.

The parameters defined in \eq{defalphas} will depend on the choice of the functions $f_1(R)$ and $f_2(R)$ and must be computed as a function of the variables for each particular model (they are analogous to the $\Upsilon$ parameter defined in Ref. \cite{Carloni:2007br}): for this, one must first invert the relation
\begin{equation}
\label{almostalpha2}
\dfrac{ f_2'(R) R }{ f_2(R)} = -\dfrac{\Om_2 y}{3(1+w)\Om_1} =  -\dfrac{  y + z \al_1(R)}{\Om_1},
\end{equation}
[where the constraint (\ref{raychaudhuridyn}) was used], in order to express the scalar curvature as a function of the dimensionless quantities defined in \eq{variables}, $R = R (y,z,\Om_1)$; one may then write the energy density as
\begin{equation}
\rho (y,z,\Om_1) = - {\ka f_1(R (y,z,\Om_1)) \over f_2(R (y,z,\Om_1))}\dfrac{\Om_1}{z},
\end{equation}
and finally compute the parameters $\al_1$ and $\al_2$, and $\al$.

As mentioned before, the dimensionality of the dynamical system (\ref{dynsystem}) can be reduced by using the two restrictions (\ref{friedmanndyn}) and (\ref{raychaudhuridyn}) stemming from the Friedmann and Raychaudhuri Eqs. (\ref{friedmann}) and (\ref{racheq}). One opts for eliminating the variables $\Om_1$ and $\Om_2$, obtaining
\begin{equation}
\begin{cases}\vspace{2mm}
\dfrac{dx}{dN}=x\left[x-y+3(1+w) \left( 1 + \dfrac{z}{y} \al_1 \right)\left(1+\dfrac{\al_2}{\al}\right) -3 w\right] \\\vspace{2mm} + 2(2+3w)(2-y)-3(1+w)z ( 1 + \al_1) + 9 (1+w) \dfrac{z}{y} \al_1 \\\vspace{2mm}
\dfrac{dy}{dN}=y\left[ 2\left(2-y\right) - \dfrac{x}{\alpha}\right]\\\vspace{2mm}
\dfrac{dz}{dN}=z\left[x\left(1-\dfrac{\alpha_1}{\al}\right)+3(1+w) \left( 1 + \dfrac{z}{y} \al_1 \right)+2\left(2-y\right)\right]
\end{cases},
\label{dynsystem4}
\end{equation}
with $\alpha_1$ defined by Eq. (\ref{defalphas}), and the eliminated variables given by the constraint (\ref{friedmanndyn}),
\begin{equation}
\Omega_1=1-x-y-z-\Omega_2,
\end{equation}
and the constraint (\ref{raychaudhuridyn}),
\begin{equation}
\Omega_2 = 3(1+w) \left( 1 + \dfrac{z}{y} \al_1(y,z,\Om_1) \right).
\end{equation}
Since $\al_1$ depends on the scalar curvature $R = R (y,z,\Om_1)$, and $\Om_1$ depends on $\Om_2$, the above is actually an implicit relation for $\Om_2 = \Om_2(x,y,z)$, which must be obtained for a given set of functions $f_1(R)$ and $f_2(R)$. As shall be detailed in the following sections, the particular models scrutinised in this study lead to straightforward simplifications of the convoluted expressions used above --- but this procedure can in principle be generalised to any choice of $f_1(R)$ and $f_2(R)$.

The determination of the fixed points of any dynamical system analysis depends crucially on the choice of the variables. The number of dynamical variables for the pure NMC case is the same as in Ref. \cite{Azizi:2014qsa}. Since in Ref. \cite{Azizi:2014qsa}, $f_1(R)=R$ and $f_2(R)$ remains unspecified, the fixed points appear as a function of $y$ (in the present notation). Different values of $y$ correspond to distinct cosmological eras, since this parameter is related to the decelerated parameter $q=1-y$ (defined in the following section) and to $w_{eff}=(2q-1)/3$. Furthermore, it also assumed $w=0$, thus limiting its scope to a Universe filled with pressureless dust.

\subsection{Physical Quantities}
With the adopted metric (\ref{metric}), the Ricci scalar reads
\begin{equation}
R=6(2H^2+\dot{H}).
\label{Ricci}
\end{equation}
One important parameter used in cosmology is the deceleration parameter
\begin{equation}
q\equiv-\dfrac{\ddot{a}a}{\dot{a}^2}=1-y,
\label{q}
\end{equation}
so that the scalar curvature may be written as
\begin{equation}
R=6H^2(1-q).
\end{equation}
Since our universe appears to be expanding at an accelerated rate, one is searching for a model with $q<0\rightarrow y > 1$. In GR this parameter yields
\begin{equation}
q=\dfrac{1}{2}(1+3w),
\label{qGR}
\end{equation}
which would require an exotic fluid with negative pressure, $w < -1/3$.

After determining the fixed points of the dynamical system for each particular choice of functions $f_1(R)$ and $f_2(R)$, one may straightforwardly determine the scale factor for each fixed point. From a direct integration of Eq. (\ref{q}) (for a fixed $y$), one obtains the general solution
\begin{equation}
a(t)=\begin{cases}\vspace{2mm}
\left({t \over t_0}\right)^{\frac{1}{2-y}},~~~~~~~~y\neq2\\\vspace{2mm}
e^{H_0 t},~~~~~~~~~~~~~~~y= 2
\end{cases}.
\end{equation}
For the first case, the scale factor evolves as a power of time, while in the second result the Hubble parameter will be constant and thus the scale factor will rise exponentially, {\it i.e.} a De Sitter phase. Note that this solution was obtained resorting (indirectly) to the definition of the Ricci scalar with the used metric.

Other important physical quantity is the energy density: one can determine its evolution for each fixed point from the continuity Eq. (\ref{conteq}). The general solution for this is the familiar result
\begin{equation}
\rho(t)=\rho_0 a(t)^{-3(1+w)}.
\label{solutionrho}
\end{equation}
Considering the definition of the variable $\Omega_2$ from Eq. (\ref{variables}), one can see that
\begin{equation}
\rho=\dfrac{\kappa f_1' \Omega_2}{f_2'\left[\Omega_2-3(1+w)\right]},
\label{Omega2}
\end{equation}
so, for a particular fixed point, it should be possible to determine the energy density from this relation. Note that for $\Omega_2=3(1+w)$ there appears to be a divergence in the density: physically, a fixed point with this value of $\Omega_2$ will correspond to a regime where $f_2'\rho\gg\kappa f_1'$.

\section{$f(R)$ Theories}\label{section:fR}
Let us now consider the case of $f(R)$ theories, in order to confirm the results obtained in Ref. \cite{Carloni:2007br}. In this case,
\begin{eqnarray}
f_1(R)=f(R)~~&,&~~f_2(R)=1 \to \\ \nonumber F=\kappa f'~~&,&~~\al_1=-\dfrac{y}{z},
\end{eqnarray}
and the constraint (\ref{raychaudhuridyn}) yields the trivial result $\Omega_2 = 0 $ [given the definition (\ref{variables}) and $f_2'(R) = 0$]; $\al$ will only depend on the derivatives of our arbitrary function $f(R)$ and $\al_2$ is not well determined, but does not appear in the equations. The dynamical system (\ref{dynsystem}) can be simplified to
\begin{equation}
\begin{cases}\vspace{2mm}
\dfrac{dx}{dN}=x(x-y)-y-3z+3w(1-x-y-z)-1\\\vspace{2mm}
\dfrac{dy}{dN}=y\left[2(2-y)-\dfrac{x}{\al} \right]\\\vspace{2mm}
\dfrac{dz}{dN}=z[2(2-y) + x ]+\dfrac{xy}{\al}
\end{cases},
\end{equation}
and the modified Friedmann Eq. yields
\begin{equation}
\Omega_1=1-x-y-z.
\end{equation}
This system is equivalent to the one presented in Ref. \cite{Carloni:2007br}, as expected (for an extensive discussion of a dynamical system approach on $f(R)$ theories see also Ref. \cite{Amendola:2006we,Carloni:2004kp}).

\section{Pure Nonminimal Coupling Case}\label{section:pureNMC}
To study the influence of the NMC in cosmology, a simple case where $f_1(R)=R$ and $f_2(R)=f(R)$ is considered. One can see that
\begin{equation}
F=\kappa-f'\rho,~~~~ F'=-f''\rho,
\end{equation}
and, from constraint (\ref{raychaudhuridyn}), $\Omega_2$ can be written as
\begin{equation}
\Omega_2=3(1+w)\left(1+\dfrac{z}{y}\right).
\label{o2restriction}
\end{equation}
Also, Eq. (\ref{alpha}) implies that $\al=\al_2(1+z/y)$, and thus the dynamical system (\ref{dynsystem}) can be written as
\begin{equation}
\footnotesize
\begin{cases}\vspace{2mm}
\dfrac{dx}{dN}=(4+x)(2+3w+x)-y[2(2+3w)+x] + \\\vspace{2mm}
~~~~~~~~~ 3(1+w)(3+x)\dfrac{z}{y} - 6(1+w)z \\\vspace{2mm}
\dfrac{dy}{dN}=y\left[2(2-y)-\dfrac{x y}{\left(y + z\right) \alpha_2}\right]\\\vspace{2mm}
\dfrac{dz}{dN}=z \bigg[2(2-y) + x + 3 \left(1 + w\right) \left(1 + \dfrac{z}{y}\right) - \dfrac{x y}{\left(y + z\right) \al_2}\bigg]
    \end{cases},
\end{equation}
while the modified Friedmann Eq. yields
\begin{equation}
\Omega_1=1-\left[x+y+z+3(1+w)\left(1 + \dfrac{z}{y}\right)\right].
\end{equation}

\subsection{Power law Nonminimal Coupling}\label{PLNC}
Let us consider a simple function
\begin{equation}
f_2(R)= C + \left(\dfrac{R}{12M^2}\right)^n,
\label{powerlaw}
\end{equation}
already studied in Ref. \cite{Bertolami:2010cw} with $C=1$. The parameters $C$ and $M$ are both constant and the latter is related to the energy scale of the theory. For this model, $\al_2=n-1$, independently of $C$. For $C=0$, the exponent $n$ should be close to zero so as to introduce a small deviation from $f_2(R)=1$; conversely, for $C=1$, $n$ may take any value. The fixed points obtained for both cases are the same, but the evolution of the physical quantities will differ. 

The fixed points of this system are obtained imposing a null variation of the dynamic variables. Their values associated with the solutions are shown in Table \ref{tablef2}.

Comparing these fixed points with the solutions obtained in the article Ref. \cite{Bertolami:2010cw}, one can verify that $\mathcal{B}$ (with $w=0$) corresponds to the $f_2'\rho\gg\kappa$ regime; $\mathcal{C}$ has some similar features to the $f_2'\rho\ll\kappa$ regime, although there is not an exact equivalence, as discussed in the corresponding paragraph. Also there appears to be an extra solution not mentioned in Ref. \cite{Bertolami:2010cw}, corresponding to a De Sitter phase of exponential expansion of the Universe.

To have a point that can replicate the effects of dark energy, one requires it to be stable with $q<0$. Also, since the model is a power law of $R$, one requires that $n<0$, so that it dominates only for late times, when the scalar curvature sufficiently small.

In the regime $f_2'\rho\gg\kappa$, one can see that
\begin{equation}
\Omega_1\sim-\dfrac{f_2}{6f_2'H^2}=-\dfrac{y}{n} \left[1 + C \left({12M^2 \over R}\right)^n \right],
\label{RelationOmega1}
\end{equation}
This relation will be used to further explore the physical significance of relevant fixed points.

\begin{widetext}

\begin{table}[!h]
\begin{footnotesize}
\caption{Fixed points and respective solutions of the model, Eq. (\ref{powerlaw}).}
\label{tablef2}
\begin{tabular}{{c}{c}{c}{c}{c}}
\hline 
\hline
 				& ($x$, $y$, $z$, $\Omega_1$,$\Omega_2$) & $a(t)$ & $\rho(t)$ & $q$  \\
\hline
$\mathcal{A}$	& $( 0, 2, 0, -4 - 3 w, 3(1+w))$&	$e^{H_0 t}$& $e^{-3(1+w)H_0t}$ & $-1$\\
$\mathcal{B}$	&$\left(\dfrac{4-2n(4+3w)}{2 n-1}, \dfrac{n ( - 2 + 4 n + 3 w )}{1 - 3 n + 2 n^2}, 0, \dfrac{2 - 4 n - 3 w}{1 - 3 n + 2 n^2},3(1+w)\right)$ &  $\left({t \over t_0}\right)^{\frac{1-3n+2n^2}{2-n(4+3w)}}$  & $\left({t \over t_0}\right)^{\frac{3(n-1)(2n-1)(1+w)}{n(4+3w)-2}}$ & $-1+\dfrac{2-n(4+3w)}{1-3n+2n^2}$ \\
$\mathcal{C}$	& $\left(\dfrac{6n (1 + w)}{1-4n - 3 w}, -\dfrac{ 1 - 4 n - 3 w}{
 2 (n-1)}, \dfrac{1 - 2 n - 3 w}{2 (n-1)}
 , \dfrac{1}{1 - n},-\dfrac{6n(1+w) }{ 1-4n-3w}\right)$ & $\left({t \over t_0}\right)^{\frac{2(1-n)}{3(1+w)}}$ & $\left({t \over t_0}\right)^{2n-2}$ & $-1+\dfrac{3(1+w)}{2(1-n)}$\\
 \hline
 \hline
\end{tabular}
\end{footnotesize}
\end{table}

\end{widetext}

\subsubsection{Point $\mathcal{A}$} \label{1pointpower}
This point corresponds to a De Sitter solution in the regime $f_2'\rho\gg\kappa$, since $\Omega_2=3(1+w)$ and considering Eq. (\ref{Omega2}). For $C=0$, the above yields $y=-n\Omega_1$, which leads to the restriction $n=2/(4+3w)$. For $C=1$, the relation (\ref{RelationOmega1}) yields
\begin{equation}
H_0=M\left[n\left(2+{3w\over 2}\right)-1\right]^{-1/(2n)},
\end{equation}
for $n > 2/(4+3w)$. The stability of the point is shown in Fig. \ref{f1Rf2Rn1}.

Notice that the NMC exponent is positive, $n>0$, so that its effect should be dominant at early times, when the curvature is high, $R \gg M^2$. Furthermore, Fig. \ref{f1Rf2Rn1} shows that the fixed point is never an atractor for any pair $(w,n)$, but unstable or a saddle point. Thus, it is not a viable candidate for dark energy, but could have some bearing on inflation.

\subsubsection{Point $\mathcal{B}$}\label{2pointpower}
\begin{figure}[h!]
  \centering
    \includegraphics[width=0.48\textwidth]{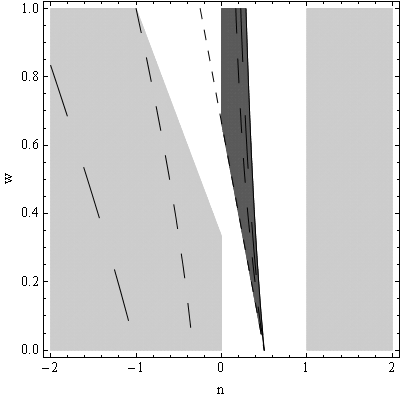}
    \caption{Stability region of $\mathcal{B}$. The light grey region corresponds to the a stable fixed point, the dark grey region to an unstable fixed point and the remaining to a saddle point. Large, medium and short dash indicate $q=0$, $q=1/2$ and $q=1$, respectively. The continuous line corresponds to $q=-1$.}
    \label{f1Rf2Rn2}
\end{figure}

Since $\Omega_2=3(1+w)$, this point is in the regime $f_2'\rho\gg\kappa$ again due to Eq. (\ref{Omega2}). If $n=2/(4+3w)$, it is equal to $\mathcal{A}$, so a De Sitter phase is attained. The stability of the point is shown in Fig. \ref{f1Rf2Rn2}.

Notice that one has $y=-\Omega_1/n$: from Eq. (\ref{RelationOmega1}), this is only physical when $C=0$ or, if $C=1$, when $(R/12M^2)^n \gg 1$ --- so that the NMC should be dominant in the latter case. This is in accordance with the corresponding regime $f_2' \rho \gg \ka$ studied in Ref. \cite{Bertolami:2010cw}, and confirmed by the value for the deceleration parameter when $w=0$, $q = (1+n)/(1-n)$, and requires a negative value for the exponent $n$ for the NMC to dominate at late times.

In GR, for the era of matter dominance, $w=0\rightarrow q=1/2$ and for the radiation era, $w=1/3\rightarrow q=1$. In Table \ref{tablepointpowerlaw} it is shown that it is possible to have quite different values of $q$ with respect to the latter: for example, it is possible to have an evolution characteristic of the radiation era in GR, i.e. $q=1$, even when $w=0$.

\begin{table}[!h]
\setlength{\tabcolsep}{9pt}
\caption{Values of $n$ needed to obtain the usual deceleration values for different $w$.}
\begin{tabular}{ccccccc}
\hline 
\hline
& $q$ & $w_{GR}$ & $w$ & $n$ & $w$ & $n$\\
\hline
$\mathcal{B}$& $-1$ & $-1$ & $0$ & $\infty$& $1/3$ & $2/5$\\
& $1/2$ & $0$ & $0$ & $-1/3$ & $1/3$ & $-1/2, 1/3$  \\
& $1$ & $1/3$ & $0$ & $0$& $1/3$ & $0, 1/4$\\
$\mathcal{C}$ & $1/2$ & $0$ & $0$ & $0$ & $1/3$ & $-1/3$  \\
& $1$ & $1/3$ & $0$ & $1/4$& $1/3$ & $0$\\
 \hline
 \hline
\end{tabular}
\label{tablepointpowerlaw}
\end{table}

\subsubsection{Point $\mathcal{C}$}\label{3pointpower}
Notice that this point has $x=-\Omega_2$ so, from Eq. (\ref{Fconstant}), $F$ is constant and its value can be determined
\begin{equation}
z=-\dfrac{\kappa R}{6FH^2}\Rightarrow F=-\dfrac{\kappa y}{z}=\kappa\dfrac{1-4n-3w}{1-2n-3w}.
\label{Fconstant3pf}
\end{equation}
For this case, one has
\begin{equation}
\rho(t)=\dfrac{24\kappa M^2}{1-2n-3w}\left[\dfrac{3^2(1+w)^2M^2}{(1-n)(1-4n-3w)}t^2\right]^{n-1},
\end{equation}
independent of $C$. For $w=0$, this reads
\begin{equation}
\rho_0=\dfrac{8}{3}\left(\dfrac{3}{4}\right)^n(1-5n+4n^2)^{-n}\dfrac{(1-n)(1-4n)}{(1-2n)}\left(\dfrac{t_0}{t_2}\right)^{2n}\dfrac{\kappa}{t_0^2},
\end{equation}
with $t_2\equiv 1 / (2\sqrt{3}M)$.

One can see that the result obtained here is different from the one attained in the $f_2' \rho \ll \rho$ regime studied in Ref. \cite{Bertolami:2010cw}: in the latter, $f_2'\rho=0$ was effectively assumed, and thus $F=\kappa$. Conversely, Eq. (\ref{Fconstant3pf}) with $w=0$ (a universe filled with pressure-less dust) reads
\begin{equation}
F=\kappa\dfrac{1-4n}{1-2n}\neq \kappa \to {f_2' \rho \over \kappa} = {2n\over 1-2n}.
\end{equation}
and indeed one finds that $f_2' \rho/\ka$ can be of the order unity or larger.

Nevertheless, both point $\mathcal{C}$ here obtained (for $w=0$) and the regime $f_2' \rho \ll \ka$ studied in Ref. \cite{Bertolami:2010cw} predict the same evolution for the scale factor, typified by a deceleration factor $q = -1+3/[2(1-n)]$.

Also, one can see that
\begin{equation}
\Omega_1=\dfrac{1}{1-n} \left[ 1 + C \left({12M^2 \over R}\right)^n \right] \to C \left({12M^2 \over R}\right)^n = 0,
\end{equation}
and a consistent solution is obtained when $C=0$, or alternatively if $C=1$ and $(R/12M^2)^n \gg 1$: the latter implies that the NMC must be dominant, again requiring a negative value for the exponent $n$ in order to replicate late time dark energy.

\begin{figure}[h!]
  \centering
    \includegraphics[width=0.48\textwidth]{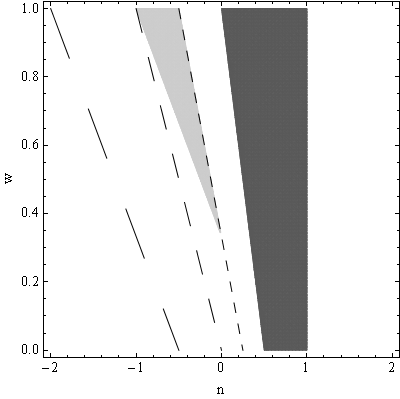}
    \caption{Stability region of $\mathcal{C}$. The light grey region corresponds to a stable fixed point, the dark grey region to an unstable fixed point and the remaining to a saddle point. Large, medium and short dash indicate $q=0$, $q=1/2$ and $q=1$, respectively.}
    \label{f1Rf2Rn3}
\end{figure}

The stability of the point is shown in Fig. \ref{f1Rf2Rn3}. $\mathcal{C}$ is not a viable candidate for dark energy, since the stable region corresponds to $q>0$. In Table \ref{tablepointpowerlaw} the values of $n$ are shown for typical values of $q$ and $w$. As expected, when $n=0$, these coincides exactly with the results of GR. Again, this fixed point allows one type of matter to mimic another ({\it e.g.} NMC dust leads to a behaviour typical of radiation in GR), as depicted on Table \ref{tablepointpowerlaw}.

\begin{figure}[h!]
  \centering
    \includegraphics[width=0.48\textwidth]{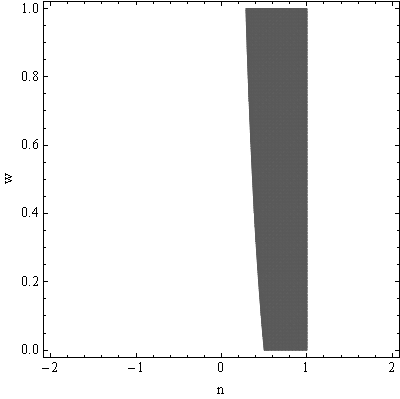}
    \caption{The dark grey region corresponds to the unstable region of $\mathcal{A}$. There is no stable region and the remaining phase space corresponds to a saddle point.}
    \label{f1Rf2Rn1}
\end{figure}

\subsection{Exponential Nonminimal Coupling}
The study of an exponential model,
\begin{equation}
f_2(R)=\exp\left(\dfrac{R}{R_0}\right),
\label{exponential}
\end{equation}
might be of interest because when the scalar curvature tends to zero, the NMC vanishes asymptotically. One can see that the effects of an exponential function in $f(R)$ theories, Ref. \cite{Abdelwahab:2007jp}, is richer than in NMC theories. For this model, one can determine $\al_2$ using Eq. (\ref{almostalpha2}), since for this particular case $\al_2=f_2''R/f_2'=f_2'R/f_2$. Also, the relation $\al=\al_2(1+z/y)$ is still valid, and thus the dynamical system is well determined. The fixed points obtained are shown in Table \ref{tablef23}.

\begin{table} [!h]
\caption{Fixed points of the model, Eq. (\ref{exponential}).}
\setlength{\tabcolsep}{6pt}
\begin{footnotesize}
\begin{tabular}{cccccc}
\hline 
\hline
	&&&&& Coordinates ($x$, $y$, $z$, $\Omega_1$,$\Omega_2$)\\
\hline
$\mathcal{A}$	&&&&& $\left(-\dfrac{3}{2} (1 + w), 2, -1, 0,\dfrac {3 }{ 2}(1+w)\right)$ \\
$\mathcal{B}$	&&&&& $(0, 2, 0, -4 - 3 w,3(1+w))$  \\
$\mathcal{C}$	&&&&& $(-4 - 3 w, 2, 0, 0,3(1+w))$  \\
 \hline
 \hline
\end{tabular}
\end{footnotesize}
\label{tablef23}
\end{table}

\subsubsection{Point $\mathcal{A}$}
This is a saddle point with no physical meaning. First, it presents the unusual case where $y=2$, so that the scalar curvature is constant, but $x\neq 0$. One can see that $x=-\Omega_2$, which implies that $F$ is constant, from Eq. (\ref{Fconstant}). From Eq. (\ref{Omega2}), one can see that $\rho=-\kappa/f_2'$. Also, considering the definitions presented in Eq. (\ref{variables}), in order to have $x\neq0$ with a constant curvature, one needs $H=0$ or $F=0$. Note that,
\begin{equation}
F=\kappa-f_2'\rho=0 \Rightarrow \rho=\kappa/f_2',
\end{equation}
which disagrees with the previous result unless $R_0 \to \infty$ and GR is recovered. For $H=0$,
\begin{equation}
\Omega_1=\dfrac{f_2\rho}{6FH^2}\rightarrow\infty\neq0,
\end{equation}
unless $\rho=0$, but it will also disagree with the previous result --- thus, proving the inconsistency of this point.

\subsubsection{Point $\mathcal{B}$}
This is a saddle point in the regime $f_2'\rho\gg\kappa$, since $\Omega_2=3(1+w)$, with
\begin{equation}
H_0^2=\dfrac{R_0}{6(4+3w)}.
\end{equation}
This is the only consistent fixed point for this model. For $y=2$, $R=12H_0^2$ is constant, implying $x=0$. Also,
\begin{equation}
z\sim\dfrac{\kappa R}{6f_2'\rho H_0^2}\sim\dfrac{\kappa}{f_2'\rho}\rightarrow 0.
\end{equation}
\subsubsection{Point $\mathcal{C}$}
This is a stable point in the regime $f_2'\rho\gg\kappa$, since $\Omega_2 = 3(1+w)$. This appears to be another case where $y=2$, $R$ constant with $x\neq0$ and it is inconsistent. The definition of $x$ from Eq. (\ref{variables}) and $f_2'\rho\gg\kappa$ implies that
\begin{equation}
x\sim-\dfrac{\dot{R}}{R_0 H}.
\end{equation}
Since $\dot{R}=0$, to obtain $x\neq0$, it is necessary that $H=0$, to induce an indetermination. When considering the definition of $\Omega_1$,
\begin{equation}
\Omega_1\sim -\dfrac{R_0}{6H^2}\rightarrow\infty\neq0,
\end{equation}
which makes this an inconsistent point. Note that this was the only point of this model with a stable region: although all the fixed points correspond to a De Sitter phase, none of them can be used to describe dark energy.

\section{Power law NMC and curvature term}\label{section:powerlaw}
Let us now consider the model 
\begin{eqnarray}
f_1(R)&=&R+12M_1^2\left(\dfrac{R}{12M_1^2}\right)^{n_1},\label{doublepowerlaw}\\
f_2(R)&=&1+\left(\dfrac{R}{12M_2^2}\right)^{n_2},\nonumber
\end{eqnarray}
where $M_i$ are characteristic energy scales. One can see that $\al_2=n_2-1$ and relation (\ref{defalphas}) is still valid; to determine $\al$, one resorts to equation (\ref{alpha}) and writes
\begin{equation}
\dfrac{f_1''R}{f_1'}=n_1\left(1-\dfrac{1}{\al_1}\right)
\end{equation}

The dynamical system is obtained by replacing all this parameters in the initial system (\ref{dynsystem4}), but it is too extensive to be presented here. The fixed points obtained from this dynamic system are shown in Table \ref{tableff} and the corresponding solutions in Table \ref{tableff2}.

Note that for $n_1=1 \rightarrow f_1(R)=2R$, the fixed points coincide with the ones presented in the subsection \ref{PLNC}, considering the restriction for $\Omega_2$, Eq. (\ref{o2restriction}). For the GR case ($n_1=n_2=0$), the fixed points obtained will collapse to only two: an unstable matter dominance and a stable cosmological constant dominance.

\setcounter{subsubsection}{0}
\subsubsection{Points $\mathcal{A}$, $\mathcal{B}$ and $\mathcal{C}$}
\begin{figure}[h!]
  \centering
    \includegraphics[width=0.46\textwidth]{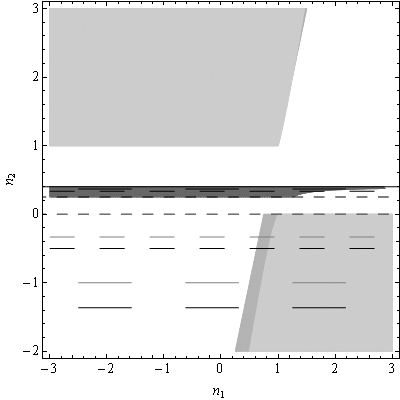}
    \caption{The two lightest grey region correspond to the stable regions of $\mathcal{B}$ when $w=1/3$ and $w=0$, from lightest to darkest, respectively. The two darkest grey regions are overlapped but correspond to an unstable region, where from the lightest to the darkest corresponds to $w=1/3$ and $w=0$. Large, medium and short dash indicate $q=0$, $q=1/2$ and $q=1$, respectively. The continuous line corresponds to $q=-1$. The black traces corresponds to $w=1/3$ and the grey to $w=0$.}
    \label{f1Rnf2Rn7}
\end{figure}

\begin{figure}[h!]
  \centering
    \includegraphics[width=0.46\textwidth]{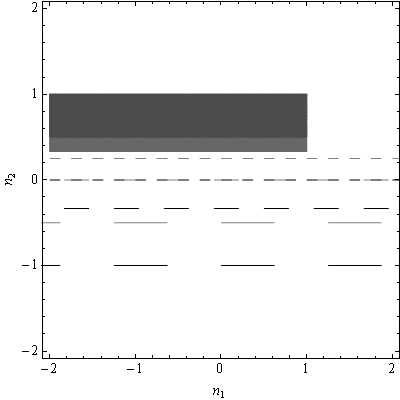}
    \caption{Stability region for point $\mathcal{C}$. There is no stable region for $w=1/3$ and $w=0$.
    The dark grey region corresponds to the unstable region when $w=0$ and lightest when $w=1/3$. The remaining regions corresponds to saddle points.  Large, medium and short dash indicate $q=0$, $q=1/2$ and $q=1$, respectively. The black traces corresponds to $w=1/3$ and the grey to $w=0$.}
    \label{f1Rnf2Rn8}
\end{figure}

These correspond exactly to the points presented in section \ref{PLNC}  of the power law pure NMC case. The stability of these points is shown in Fig. \ref{f1Rf2Rn1},\ref{f1Rnf2Rn7} and \ref{f1Rnf2Rn8}, respectively. Despite this correspondence between the points of both models, the stability of $\mathcal{B}$ and $\mathcal{C}$ is altered by the NMC.

\subsubsection{Points $\mathcal{D}$, $\mathcal{E}$ and $\mathcal{F}$}
\begin{figure}[h!]
  \centering
    \includegraphics[width=0.46\textwidth]{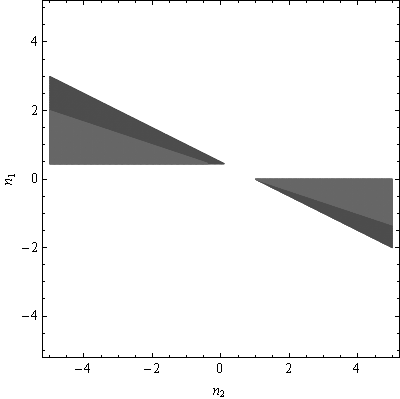}
    \caption{Stability region for $\mathcal{E}$. There is no stable region for $w=1/3$ and $w=0$. The unstable region for $w=0$ (dark grey) is overlapped with the $w=1$ (light grey). The remaining regions corresponds to saddle points.}
    \label{f1Rnf2Rn9}
\end{figure}

These are points of little interest with an evolution similar to the radiation era. Since $\Omega_2=0$, one might expect these points to be related to pure $f(R)$ solutions. In fact, they appear in Ref. \cite{Carloni:2007br}, that studies the $f(R)=R+R_0(R/R_0)^n$ model.
$\mathcal{D}$ is a saddle point, the stability of $\mathcal{E}$ is shown in Fig. \ref{f1Rnf2Rn9} and $\mathcal{F}$ is a saddle point when $0<w<2/3$, and for $w >2/3$ the stability depends on both parameters $n_1$ and $n_2$.

\subsubsection{Point $\mathcal{G}$}
\begin{figure}[h!]
  \centering
    \includegraphics[width=0.46\textwidth]{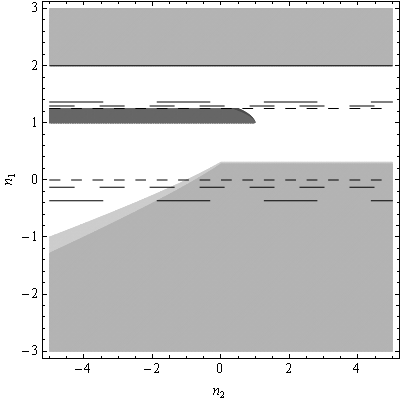}
    \caption{The two lightest grey region correspond to the stable regions of $\mathcal{G}$ when $w=1/3$ and $w=0$, from lightest to darkest respectively. For $n_1>2$, the regions overlap and are both stable. The two darkest grey regions are overlapped but correspond to an unstable region for $w=1/3$ and $w=0$.  Large, medium and short dash indicate $q=0$, $q=1/2$ and $q=1$, respectively. The continuous line corresponds to $q=-1$.}
    \label{f1Rnf2Rn4}
\end{figure}

In this point the energy density is null so the NMC is neglected, $f_2'\rho\ll\kappa f_1'$. This corresponds to a point based on a pure $f(R)$ theory that was studied in Ref. \cite{Carloni:2007br}. The stability of the point is shown in Fig. \ref{f1Rnf2Rn4}.

It is easy to see that a De Sitter phase is obtained when $n_1=2$, which corresponds to $x=0$, $y=2$ and $z=-1$ --- the same as the previous fixed point, but their origin is completely different. The Starobinsky inflation model, Ref. \cite{Starobinsky:1980te}, corresponds to $n_1=2$ and $n_2=0$. Since the NMC is neglected, this point corresponds to that solution.

The scale factor and deceleration parameter are independent of $w$, but the stability has some dependency, as shown in Fig. \ref{f1Rnf2Rn4}. One can see that when
\begin{equation}
n_1=(7\pm\sqrt{73})/12 \rightarrow q=1/2,
\end{equation}
and
\begin{equation}
n_1=0, 5/4 \rightarrow q=1.
\end{equation}

\subsubsection{Point $\mathcal{H}$}
\begin{figure}[h!]
  \centering
    \includegraphics[width=0.46\textwidth]{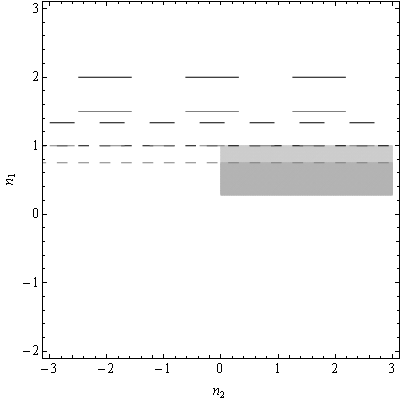}
    \caption{The two overlapped regions correspond to the stable regions of $\mathcal{H}$ when $w=1/3$, light grey, and $w=0$, dark grey. There is no unstable region for both cases. The remaining region corresponds to saddle points. Large, medium and short dash indicate $q=0$, $q=1/2$ and $q=1$, respectively. Black traces correspond to $w=1/3$ and grey traces to $w=0$.}
    \label{f1Rnf2Rn5}
\end{figure}

The scale factor only depends on $n_1$, reflecting a stronger influence of $\kappa f_1'$. This is also visible, since $\Omega_1\neq0$ and $\Omega_2=0$. The regime $f_2'\rho\ll\kappa f_1'$ is verified, but with non-null energy density. This fixed point also appears in Ref. \cite{Carloni:2007br}, and is based on pure $f(R)$ theory. The stability of the point is shown in Fig. \ref{f1Rnf2Rn5}.

A De Sitter solution is only obtained when $w=-1$, which is similar to the use of a cosmological constant. Table \ref{tablepoint5} shows the values of $n_1$ needed to obtain the usual values of $q$, when $w=0, 1/3$. For $n_1=1$, the result $q=(1+3w)/2$, typical of GR, is recovered; interestingly, this does not depend on $n_2$.

\begin{table}[!h]
\caption{Values of $n$ needed to obtain the usual deceleration values for different $w$ for $\mathcal{H}$.}
\setlength{\tabcolsep}{11pt}
\begin{tabular}{cccccc}
\hline 
\hline
$q$ & $w_{GR}$ & $w$ & $n_1$ & $w$ & $n_1$\\
\hline
$1/2$ & $0$ & $0$ & $1$ & $1/3$ & $4/3$  \\
$1$ & $1/3$ & $0$ & $3/4$& $1/3$ & $1$\\
 \hline
 \hline
\end{tabular}
\label{tablepoint5}
\end{table}

\subsubsection{Point $\mathcal{I}$}
This is stable when $1<n_1<2$ and a saddle point in rest of the region. It corresponds to a De Sitter phase, with no matter. The Hubble parameter has to satisfy the condition
\begin{equation}
H_0^2=M_1^2\left(\dfrac{1}{n_1-2}\right)^{1/(n_1-1)},
\end{equation}
for $n_1\neq1$ and $n_1\neq2$. This point also appears in Ref. \cite{Carloni:2007br}.

\subsubsection{Point $\mathcal{J}$}
\begin{figure}[h!]
  \centering
    \includegraphics[width=0.46\textwidth]{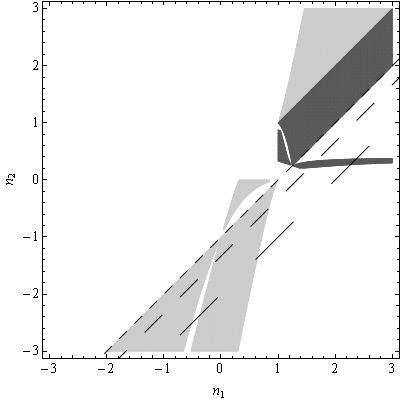}
    \caption{Stability region of $\mathcal{J}$ when $w=1/3$. Light grey corresponds to the stable region and the dark grey to the unstable. The remaining corresponds to a saddle point. Large, medium and short dash indicate $q=0$, $q=1/2$ and $q=1$, respectively.}
    \label{f1Rnf2Rn613}
\end{figure}

\begin{figure}[h!]
  \centering
    \includegraphics[width=0.46\textwidth]{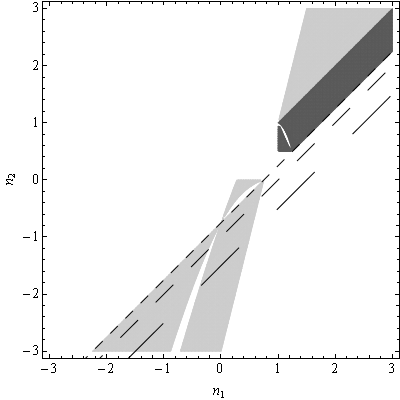}
    \caption{Stability region of $\mathcal{J}$ when $w=0$. Light grey corresponds to the stable region and the dark grey to the unstable. The remaining corresponds to a saddle point. Large, medium and short dash indicate $q=0$, $q=1/2$ and $q=1$, respectively.}
    \label{f1Rnf2Rn60}
\end{figure}

This point appears to be a generalization of $\mathcal{C}$ and $\mathcal{H}$, has no dominant regime and it is the only point that depends explicitly on both functions.  The stability of the point is shown in Fig. \ref{f1Rnf2Rn613} and \ref{f1Rnf2Rn60}, for $w=1/3$ and $w=0$, respectively. This point is divergent when $n_1=n_2$.

A De Sitter solution is only obtained when $w=-1$, which is similar to the use of a cosmological constant. Table \ref{tablepoint6} shows the relation between $n_1$ and $n_2$ needed to obtain the usual values of $q$, when $w=0, 1/3$. The normal results for GR are obtained when $n_1-n_2=1$.

\begin{table}[!h]
\caption{Values of $n$ needed to obtain the usual deceleration values for different $w$ for $\mathcal{J}$.}
\setlength{\tabcolsep}{6pt}
\begin{tabular}{cccccc}
\hline 
\hline
$q$ & $w_{GR}$ & $w$ &  & $w$ & \\
\hline
$1/2$ & $0$ & $0$ & $n_1-n_2=1$ & $1/3$ & $n_1-n_2=4/3$  \\
$1$ & $1/3$ & $0$ & $n_1-n_2=3/4$& $1/3$ & $n_1-n_2=1$\\
 \hline
 \hline
\end{tabular}
\label{tablepoint6}
\end{table}

\subsection{Modified Friedmann equation}
In this section we attempt a comparison between the results obtained above and those of Ref. \cite{Bertolami:2013uwl}, which is based upon the phenomenological study of modifications of the Friedmann equation, of the form
\begin{equation}
H^2\sim\rho^{1+\beta}.
\end{equation}
As shown in that study, the above relation can be obtained in the regime $F=\mbox{const.}$ when $f_1(R)\sim R$ and $f_2(R)\sim R^{\beta/(1+\beta)}$. From Eq. (\ref{solutionrho}), the scale factor is given by
\begin{equation}
a(t)\sim t^{\frac{2}{3(1+w)(1+\beta)}},
\end{equation}
and the deceleration parameter is
\begin{equation}
q=-1+\dfrac{3(1+\beta)(1+w)}{2}.
\label{qcompare}
\end{equation}
From the above, a comparison between this solutions and the solutions obtained from the fixed points is possible.

Considering the exponents of the functions defined in Eq. (\ref{doublepowerlaw}) as $n_1=1$ and $n_2=\beta/(1+\beta)$, one can see that $\mathcal{C}$, $\mathcal{H}$, $\mathcal{I}$ and $\mathcal{J}$ correspond to a constant $F$. $\mathcal{I}$ corresponds to a De Sitter phase and $q$ is only equal to Eq. (\ref{qcompare}) when $\beta=-1$, which corresponds to $n_2\rightarrow\infty$. $\mathcal{C}$ and $\mathcal{J}$ have a deceleration parameter exactly like Eq. (\ref{qcompare}). $\mathcal{B}$ has $F=\mbox{const.}$ and satisfies Eq. (\ref{qcompare}) when $\beta=(1-3w)/(1+3w)$. $\mathcal{H}$ also satisfies Eq. \ref{qcompare} when $\beta=0$, which corresponds to GR case. Thus, one concludes that the modifications to Friedmann equation due to a NMC are indeed obtainable from a dynamical system's approach, as correctly argued in Ref. \cite{Bertolami:2013uwl}.

\begin{widetext}

\begin{table}[!h]%
\caption{Fixed points of the model, Eq. (\ref{doublepowerlaw}).}
\begin{footnotesize}
\begin{tabular}{lp{10pt}lp{5pt}lp{5pt}lp{5pt}lp{5pt}lp{5pt}r}
\hline 
\hline
Point 				&& Coordinates ($x$, $y$, $z$, $\Omega_1$, $\Omega_2$) \\
\hline
$\mathcal{A}$	&& $ (0, 2, 0, -4 - 3 w, 3 (1 + w)) $ \\
$\mathcal{B}$	&& $\left(\dfrac{4 - 2 n_2 (4 + 3 w)}{-1 + 2 n_2}, \dfrac{n_2 (-2 + 4 n_2 + 3 w)}{1 - 3 n_2 + 2 n_2^2}, 0, \dfrac{2 - 4 n2 - 3 w}{1 - 3 n2 + 2 n2^2},  3 (1 + w)\right)$ \\
$\mathcal{C}$	&& $\left( -\dfrac{6 n_2 (1 + w)}{-1 + 4 n_2 + 3 w},\dfrac{1 - 4 n_2 - 3 w}{2 - 2 n_2}, \dfrac{1 - 2 n_2 - 3 w}{2 (-1 + n_2)},\dfrac{1}{1 - n_2}, \dfrac{6 n_2 (1 + w)}{-1 + 4 n_2 + 3 w}\right)$ \\
$\mathcal{D}$	&& $ ( -4, 0, 5, 0, 0) $ \\
$\mathcal{E}$ && $\left(1, 0, 0, 0, 0\right)$\\
$\mathcal{F}$ && $\left(-1+3w, 0, 0, 2-3w, 0\right)$\\
$\mathcal{G}$	&& $ \left(-\dfrac{2 (n_1-2)}{2 n_1-1}, \dfrac{n_1 (-5 + 4 n_1)}{1 - 3 n_1 + 2 n_1^2}, \dfrac{5 - 4 n_1}{1 - 3 n_1 + 2 n_1^2}, 0, 0\right) $ \\
$\mathcal{H}$	&& $\left(\dfrac{3 (-1 + n_1) (1 + w)}{n_1}, -\dfrac{3 - 4 n_1 + 3 w}{2 n_1}, \dfrac{3 - 4 n_1 + 3 w}{2 n_1^2}, \dfrac{-3 (1 + w) - 2 n_1^2 (4 + 3 w) + n_1 (13 + 9 w)}{2 n_1^2}, 0\right)$ \\
$\mathcal{I}$	&& $ (0, 2, -1, 0, 0) $ \\
\multirow{3}*{$\mathcal{J}$}	&& $\bigg(-\dfrac{3 (1 + w) \left[(-1 + n_2) (3 + 4 n_2 + 3 w) + n_1 \left[7 - 2 n_2^2 + 3 w - 9 n_2 (1 + w)\right] + n_1^2 \left[-4 + n_2 (8 + 6 w)\right]\right]}{(n_1 - n_2) [4 n_1 - 4 n_2 - 3 (1 + w)]},$\\
&& $-\dfrac{3 - 4 n_1 + 4 n_2 + 3 w}{2 n_1 - 2 n_2},\dfrac{3 - 5 n_2 - 2 n_2^2 + 3 w - 9 n_2 w + n_1 [-4 + n_2 (8 + 6 w)]}{2 (n_1 - n_2)^2},$\\
&& $\dfrac{-4 n_2 - 3 (1 + w) - 2 n_1^2 (4 + 3 w) + n_1 (13 + 2 n_2 + 9 w)}{2 (n_1 - n_2)^2},\dfrac{3 n_2 (1 + w) [3 + 4 n_2 + 3 w + n_1^2 (8 + 6 w) - n_1 (13 + 2 n_2 + 9 w)]}{(n_1 - n_2) [4 n_1 - 4 n_2 - 3 (1 + w)]}\bigg)$ \\
 \hline
 \hline
\end{tabular}
\end{footnotesize}
\label{tableff}
\end{table}

\begin{table}[!h]%
\caption{Solutions associated with the fixed points of the model, Eq. (\ref{doublepowerlaw}).}
\begin{center}
\begin{tabular}{lp{10pt}lp{5pt}lp{5pt}lp{5pt}lp{5pt}lp{5pt}r}
\hline 
\hline
Point 				&& $a(t)$ && $\rho(t)$ && $q$\\
\hline
$\mathcal{A}$	&& $e^{H_0 t}$ && $e^{-3(1+w)H_0t}$ && $-1$\\
$\mathcal{B}$	&& $\left({t \over t_0}\right)^{\frac{1-3n_2+2n_2^2}{2-4n_2-3n_2w}},~~n_2\neq\dfrac{2}{4+3w}$ && $\left({t \over t_0}\right)^{\frac{3(n-1)(2n-1)(1+w)}{n(4+3w)-2}}$ && $-1+\dfrac{2-n_2(4+3w)}{1-3n_2+2n_2^2}$\\
$\mathcal{C}$	&& $\left({t \over t_0}\right)^{\frac{2(1-n_2)}{3(1+w)}}$ && $\left({t \over t_0}\right)^{2(n_2-1)}$&& $-1+\dfrac{3(1+w)}{2(1-n_2)}$\\
$\mathcal{D}$	&& $\sqrt{{t /t_0}} $ && $0$ && $1$\\
$\mathcal{E}$ && $\sqrt{{t /t_0}} $ && $0$ && $1$\\
$\mathcal{F}$ && $\sqrt{{t /t_0}} $ && $\left({t \over t_0}\right)^{-3(1+w)/2}$ && $1$\\
$\mathcal{G}$	&& $\begin{cases}e^{H_0 t},~~~~~~~~~~~~~~~~~~~n_1=2\\\left({t \over t_0}\right)^{\frac{-1+3n_1-2n_1^2}{n_1-2}},~~~~n_1\neq2\end{cases}$ && $0$ && $-1+\dfrac{2-n_1}{1-3n_1+2n_1^2}$\\
$\mathcal{H}$	&& $\left({t \over t_0}\right)^{\dfrac{2n_1}{3(1+w)}}$ && $\left({t \over t_0}\right)^{-2n_1}$ && $-1+\dfrac{3(1+w)}{2n_1}$\\
$\mathcal{I}$	&& $e^{H_0 t}$ && $0$ && $-1$\\
$\mathcal{J}$	&& $\left({t \over t_0}\right)^{\frac{2(n_1-n_2)}{3(1+w)}}$ && $\left({t \over t_0}\right)^{2(n_2-n_1)}$ && $-1+\dfrac{3(1+w)}{2(n_1-n_2)}$\\
 \hline
 \hline
\end{tabular}
\end{center}
\label{tableff2}
\end{table}

\end{widetext}

\subsection{Linear NMC}
Note that, for $n_2=1$ divergences appear in some points and a more detail study by direct substitution is required, which is done in Ref. \cite{Azizi:2014qsa}. A direct comparison between the points obtained above and the ones attained in Ref. \cite{Azizi:2014qsa} is not done due to the different choice of the variables: in particular, note that that study resorts to a variable proportional to $\rho^2$.

Nevertheless, one can compare the deceleration parameter obtained in both works ($n_1=n$), which clearly marks the physical significance of the underlying fixed points: one finds that $\mathcal{J}$ presented here has the same deceleration parameter, $q=(5-2n)/(2(n-1))$, of the fourth point of the mentioned article. Also, there is a fixed point in Ref. \cite{Azizi:2014qsa} with $q=-1$, that can be related to the fixed points obtained here with the same value. These comparisons are only valid when the power law term dominates over $R$, since our choice of model was Eq. (\ref{doublepowerlaw}).

\section{Discussion and Outlook}\label{section:conclusion}
In this work, a dynamical system approach was made on NMC theories. The dynamical system for the most general case with two arbitrary functions was obtained. Also, the solutions and their stability for three different models were obtained and compared with previous works.

As expected, the NMC dynamical system can be particularized to a pure $f(R)$ theory when $f_2(R)=1$, yielding the same results obtained in Ref. \cite{Carloni:2007br}. One can see that the variable $\Omega_2$ introduced by the NMC, is the key to determine whether $F$ is constant or if the NMC dominates over the usual $f(R)$ theory.

In the pure NMC case described by a power law, the solutions obtained are in agreement with the ones presented in Ref. \cite{Bertolami:2010cw}. In addition, the obtained result for the energy density for point $\mathcal{C}$ is different from the one in Ref. \cite{Bertolami:2010cw}, due to the assumption of the latter that $\rho=0\rightarrow F=\kappa$, which differs from the result here obtained, Eq. (\ref{Fconstant3pf}).

Furthermore, the pure NMC exponential case appears to have less diversity of solutions than the usual exponential $f(R)$ model, as seen in Ref. \cite{Abdelwahab:2007jp}.

The last model considered of power law corrections to GR yields the solutions for the pure $f(R)$ case and the pure NMC case, as if it considered the regimes for which function dominates over the other, and also a solution that depends simultaneously on both models. Furthermore, it was determined which fixed points correspond to the general solution of $H^2\sim\rho^{1+\beta}$, presented in Ref. \cite{Bertolami:2013uwl}.

This method is a good way to determine the solutions of a particular model, since it does not assume solutions a priori. Note that, in the pure NMC power law case, there is a solution obtained by this method not considered in Ref. \cite{Bertolami:2010cw}.

However, and despite its success in determining a variety of solutions, this method depends on the chosen variables --- and can thus present a hidden selection bias, by excluding other interesting regimes not represented with the adopted set.

In addition, the existence of fixed points with the desired local stability does not imply that there is a trajectory in the phase space (i.e. a history for the Universe) that connects these points, as noted in Ref. \cite{Carloni:2007br}: ideally,  the fixed points yielding the current phase of accelerated expansion should have a extremely wide basin of attraction (or infinite, {\it i.e.} a global attractor), so that the matter dominance epoch evolves towards the former with no strong dependence on the initial conditions --- thus excluding the need for unphysical fine-tuning.

Such endeavour, which is beyond the scope of this work, could also encompass an accurate numerical analysis of the field equations, in order to estimate reasonable physical parameters ({\it e.g.} $n$ and $M$ for a power-law NMC) compatible with current cosmographic results, and derive predictions for the future evolution of cosmological parameters.

\section*{Acknowledgements}
The authors thank O. Bertolami for fruitful discussion, and the referee for his/her useful remarks. J.P. is partially supported by Funda\c{c}\~ao para a Ci\^encia e Tecnologia under the project PTDC/FIS/111362/2009. 

\end{document}